# Nonreciprocal Transmission of Microwave Acoustic Waves in Nonlinear Parity-Time Symmetric Resonators

Linbo Shao[1,*], Wenbo Mao[1,2], Smarak Maity[1], Neil Sinclair[1,3], Yaowen Hu[1], Lan Yang[2], and Marko Lončar[1,*]

[1]John A. Paulson School of Engineering and Applied Sciences, Harvard University, 29 Oxford Street, Cambridge, MA 02138, USA
[2]Department of Electrical and Systems Engineering, Washington University, St. Louis, MO 63130, USA
[3]Division of Physics, Mathematics and Astronomy, and Alliance for Quantum Technologies (AQT), California Institute of Technology, 1200 E. California Boulevard, Pasadena, CA 91125, USA
*Correspondence to: shaolb@seas.harvard.edu (L.S.); loncar@seas.harvard.edu (M.L.)

**Abstract**

Acoustic waves have emerged as versatile on-chip information carriers with applications ranging from microwave filters to transducers. Nonreciprocal devices are desirable for the control and routing of high-frequency phonons. This is challenging, however, due to the linear response of most acoustic systems. Here, we leverage the strong piezoelectricity of lithium niobate to demonstrate fully tunable gain, loss, and nonlinearity for surface acoustic waves using electric circuitry. This allows the construction of a nonlinear acoustic parity-time-symmetric system and enables nonreciprocal transmission. We achieve a nonreciprocity of 10 decibels for a 200-MHz acoustic wave at a low input power of 3 µW and further demonstrate one-way circulation of acoustic waves by cascading nonreciprocal devices. Our work illustrates the potential of this piezoelectric platform for on-chip phononic processing and exploration of non-Hermitian physics.

**Main text**

The reciprocity theorem, which generally applies to linear time-invariant systems, states that the relative field amplitude and phase at a source and a detector will be identical when their positions are swapped. Nonreciprocal or one-way acoustic transmission can be achieved by breaking the time-reversal symmetry of propagating waves. Recent advances in acoustics[1-6] feature microwave frequencies on integrated solid-state platforms, in which continuously operating, non-magnetic acoustic nonreciprocal devices on a compact platform are desired for phonon isolation and networking. Most demonstrations of asymmetric acoustic transmission have been realized in bulk media utilizing circulating fluids[7], superlattices with nonlinear media[8,9], macroscopic metamaterials[10,11], the viscosity of water[12], and the acoustic radiation pressure effect[13], but are all limited to operating frequencies below a few megahertz. Meanwhile, ferromagnetic materials have been used to demonstrate nonreciprocity at microwave frequencies, but only with weak isolation[14,15]. Moreover, the usage of ferromagnetic devices is limited by the incompatiblility of magnetic fields with many solid-state systems including superconductors and electron spins. A recent proposal using optomechanics[16] provides an alternative solution for nonreciprocal phonon transmission, but is experimentally challenging and has yet to be demonstrated.

Parity-time(PT)-symmetric systems[17-20] are formed by coupled non-Hermitian subsystems with balanced gain and loss or equivalent configurations, and have been explored in integrated optical[21-28], acoustic[29-35], and electrical[36] platforms for their rich physics and unique functionalities for practical applications. In this work, we construct a PT-symmetric system of surface acoustic waves (SAWs) at frequencies of hundreds of megahertz. SAWs are elastic waves that propagate along the surface of solid materials. They have been widely used for signal processing[37] and recently proposed as a universal interface for solid-state quantum systems such as superconducting qubits[1,3], defect centers[4,6], and nanomechanical oscillators[5]. To introduce acoustic gain and nonlinearity, we take advantage of the fact that SAWs in piezoelectric materials are not pure elastic waves but also have electric field components. By leveraging the strong piezoelectricity of lithium niobate (LN), we demonstrate an electrically tunable gain for SAWs that can fully compensate loss in a SAW resonator, as evidenced by self-oscillation. Moreover, the nonlinearity in our SAW amplifier allows nonreciprocal transmission in our PT-symmetric system in the broken symmetry regime. Along with the established electro- and acousto-optic properties of LN, our approach for acoustic control renders LN as an attractive solid-state platform to realize integrated and hybrid signal processing as well as fully controlled systems to study PT symmetry and non-Hermitian physics.

Our PT-symmetric SAW system consists of two coupled SAW resonators, defined by Bragg mirrors. Each resonator contains distributed cross-finger electrodes, known as interdigital transducers (IDTs), that are connected to external electronic circuits to provide gain and loss (Figs. 1**a** and 1**b**). SAWs propagate along the crystalline X direction on the surface of 128° Y-cut LN, and this configuration provides strong electromechanical coupling and low propagation loss[38]. The Bragg mirrors feature >30 dB reflectivity at a frequency of 200 MHz, over a bandwidth of 8 MHz (Fig. 1**c**), and are fabricated by etching grooves on LN with a pitch of 10 μm to match the acoustic half-wavelength[39]. Within the reflection band of the Bragg mirror, a SAW resonator without the IDT exhibits three resonant modes with intrinsic quality factors of up to $10^4$. Two more IDTs, which are situated outside of the coupled resonator system, are used as an emitter-receiver pair for transmission measurements. These IDTs are designed to have bandwidths larger than the Bragg reflection band, and their transmission of -14 dB (Fig. 1**c**) is excluded in the characterization of our PT-symmetric resonators. We refer to the resonator with gain (loss) as the active (passive) resonator, and we consider the forward direction of propagation to be from the passive to the active resonator (Fig. 1**a**).

In a PT-symmetric system, there are two distinct regions of the parameter space, determined by a coupling strength $\mu$ between the resonators, and their gain $g > 0$ and loss $\gamma < 0$[34]. In the unbroken PT-symmetric regime, $\mu > (g + |\gamma|)/2$. The real parts of the eigenvalues of the system are different, resulting in a mode splitting of the observed transmission spectrum (Fig. 1**c**). The transmission is reciprocal in this regime. In the broken-PT-symmetric regime, the coupling between two resonators is weak, $\mu < (g + |\gamma|)/2$, and the eigenvalues share identical real parts. This corresponds to a single resonance peak in the transmission spectrum, as we observe experimentally in Fig. 1c. In the presence of additional nonlinearities, the unequal localization in the active resonator between the forward and backward inputs can result in nonreciprocal transmission[22,31]. Details of our acoustic PT-symmetric system are discussed in Sec. II of Supplementary Material.

We numerically evaluate the nonreciprocal response of the coupled resonator system in the case of an 80-groove Bragg mirror between them (Figs. 1d and 1e). In this case, the weak coupling between the resonators puts the system in the broken PT-symmetric regime. The energy flow between the two resonators is too small to reach equilibrium. As a result, the forward propagating wave (Fig. 1d) is first attenuated in the passive resonator, before entering the active resonator featuring nonlinear gain. Thus, the wave experiences the high-gain regime of the nonlinear active resonator, the signal is amplified, and the overall system transmission is high. On the other hand, the backward propagating wave (Fig. 1e) first enters the nonlinear active resonator. Due to its high amplitude, the wave experiences lower gain due to gain saturation. Next, it enters the passive resonator, where it experiences loss before it is detected via the IDT, resulting in a low transmission. The PT-symmetry breaking induces a stronger localization in the active resonator for the backward propagating wave than that of the forward wave. The stronger localization results in a lower gain and leads to a nonreciprocal transmission. On the other hand, the use of a 30-groove Bragg mirror between the two resonators yields stronger coupling than that of the 80-groove mirror case, putting the system in the unbroken PT-symmetric regime. Here, an equilibrium between the two resonators is reached, resulting in similar SAW amplitudes in the active and passive resonators. Thus, the transmission spectra for backward and forward propagating SAWs are reciprocal (see Sec. III in Supplementary Material).

In our experiment, we first use a single resonator geometry to characterize the SAW gain and nonlinearity from an additional IDT inserted in the middle of the resonator and connected to a variable-gain electronic circuit (Fig. 2a). Specifically, the IDT is connected (using wire bonding) to a negative resistance electric circuit implemented by an operational amplifier with feedback resistors (Fig. 2b). The trimmer resistor $R_f$ in Fig. 2b tunes the effective negative resistance of the circuit (see Sec. V in Supplementary Material). Instead of dissipating electric energy, a negative resistor circuit outputs electric energy with applied voltage, and thus it provides the gain for SAWs (via the piezoelectric effect). Due to the presence of gain and noise in the resonator, a SAW can be detected via the receiving IDT without input. Below a certain gain threshold, the spectrum of the output SAW features a resonance linewidth of 94(1) kHz, which is determined by the loss of the resonator. Detailed transmission measurements are performed in this regime to characterize the SAW amplification process (Fig. S11). When the gain is increased, SAW self-oscillation occurs, and the linewidth of the output SAW narrows significantly (Fig. 2c). This indicates that the gain can fully compensate for the loss of the SAW resonator. Importantly, our negative resistance circuit features nonlinear gain: for high input signal, the operational amplifier operates in the saturation regime, in which its output voltage is clamped by the power supply voltages. The gain saturation is indicated by the reduced transmission peaks and the increased linewidths at higher input power (Figs. 2**d** and 2**e**). Similar to optical systems[22,24,40,41], this saturation nonlinearity breaks the time-reversal symmetry for SAWs, but not the dynamic reciprocity[42]. Meanwhile, SAW loss is implemented by an IDT connected to a resistor, and it behaves linearly within our range of input powers (Fig. S10).

Our PT-symmetric SAW devices can operate in the unbroken or broken symmetry regimes by adjusting the coupling strength, gain, or loss of the resonators (see Sec. I in Supplementary Material). The coupling strength is experimentally controlled by varying the number of Bragg mirror grooves between the two resonators and varies from $\mu$ = 295 kHz for 20 grooves to $\mu$ = 9 kHz for 90 grooves (Fig. S12). An 80-groove mirror ($\mu$ =27 kHz) is used for measurements in the nonreciprocal broken PT-symmetric regime; while a 30-groove mirror ($\mu$ = 180 kHz) is used for the reciprocal unbroken PT-symmetric regime.

Nonreciprocal transmission is observed in the broken PT-symmetric regime (Fig. 3**a**), in which the loss and gain of the resonators are tuned to maximize the nonreciprocal isolation. At the optimum input power of 3 µW (-25 dBm), a Lorentzian resonance peak is observed in the forward transmission spectrum, while a reduced transmission near the SAW resonance frequency is observed in the backward transmission spectrum. A nonreciprocity of $\eta$ =10.9 dB is calculated from $\eta = \frac{S_{\text{Fwd}}(f_0)}{S_{\text{Bwd}}(f_0)}\Big|_{S_{\text{Fwd}}(f_0)=\text{Max}(S_{\text{Fwd}})}$, where $S_{\text{Fwd}}(f)$ and $S_{\text{Bwd}}(f)$ represent the power transmission coefficients in the forward and backward directions, respectively, and the frequency $f_0$ corresponds to the maximum forward transmission. The forward and backward transmission spectra are measured separately by launching a SAW from one input at a time. The measured maximum forward power transmission of $S_{\text{Fwd}}$ = 0.2 corresponds to an insertion loss of 7 dB. The full-width-at-half-maximum bandwidth of the forward input is measured to be 28 kHz, which is consistent with the linewidth of the active SAW resonator. The nonreciprocity varies with power because the nonlinearity is induced by saturation (Fig. S13). The dynamic range of the input SAW, which is defined by a 3-dB degradation of nonreciprocity, is measured to be 8 dB, ranging from -28 dBm (1.6 µW) to -20 dBm (10 µW). At a lower input power of 250 nW, SAWs propagating in both directions experience linear gain, and thus no significant nonreciprocity, while at a high input power of 40 µW, the nonreciprocity is reduced since SAWs propagating in both directions experience strong nonlinear saturation.

Our experiments clarify that the nonreciprocity of the nonlinear PT system originates from the inequality of nonlinear gains induced by PT symmetry breaking for the forward and backward inputs, and eliminate the confusion about the origin of nonreciprocity in optical PT-symmetric systems[22], in which full control and direct observation of nonlinearity are not available. Furthermore, the desired SAW transmission in the forward direction can be linear and free from nonlinear distortion. Under the optimal input power of 3 µW, our PT-symmetric system remains in the linear regime for forward-propagating signals. However, for backward-propagating signals, stronger localization in the active resonator triggers strong nonlinear saturation and suppresses the transmission. We use the relative powers at higher harmonics to characterize the nonlinearity of the transmitted signals (Fig. S14). For forward transmission, the power ratio of the second harmonic is only -31 dBc (0.08%) at the optimal input power. For backward transmission, at the same input power, the power ratio of the second harmonic is as high as -13 dBc (5%).

In contrast, reciprocal transmission is observed in the unbroken PT-symmetric regime (Fig. 3**b**) using the resonator with 30-groove mirror (coupling strength of $\mu$ = 180 kHz) and the same nonlinear gain and loss circuit. At the same input power of 3 µW, similar transmission spectra with split resonances are observed for both forward and backward SAW inputs. Though saturation nonlinearity is observed with increasing input powers in both transmission directions, only reciprocal transmission is observed in this unbroken regime.

The concept of non-reciprocal SAW transmission can be leveraged in many applications. As an example, we demonstrate one-way circulation of SAWs using two nonreciprocal devices (Figs. 4**a** and 4**b**). Circulation is observed at the resonance frequency of the device. A SAW from Port 1 is almost entirely transmitted to Port 2 ($S_{21}$). At the frequency of the maximum transmission in the $S_{21}$ spectrum, the transmission in the clockwise direction $S_{21}$ is 20 dB higher than that in the counterclockwise direction $S_{31}$ (Fig. 4**c**). Similarly, for a SAW from Port 2, the transmission to Port 3 is 10 dB higher than that to Port 1 (Fig. 4**d**). The higher off-resonance transmission between Port 3 and Port 1 is caused by electrical cross-talk due to the proximity of the IDTs. The discrepancy between different input ports is likely caused by the small variation in resonance frequencies and *Q* factors of the acoustic resonators.

The performance of our nonreciprocal SAW devices is not fundamentally limited and could be improved. We evaluate the dependence of the nonreciprocity, the dynamic range, and the insertion on the device parameters including the coupling strengths and the gain in Sec. III of Supplementary Material. From the theoretical analysis, a nonreciprocity of over 35 dB, a dynamic range of over 50 dB, and an insertion loss of 0 dB could each be feasible with our nonreciprocal SAW resonators. However, there is a trade-off between various parameters. Generally, a weaker coupling strength between the SAW resonators leads to a larger dynamic range, but greater insertion loss; higher gain of the active resonator will result in stronger nonreciprocity, lower insertion loss, but a smaller dynamic range. In practice, the gain and nonlinearity are limited by the commercially available operational amplifiers used in our demonstration, which are not designed to have strong nonlinearities. An integrated circuit designed specifically for this purpose or a nonlinear superconducting circuit could further improve the nonreciprocity of our SAW devices. We note that use of the gain would inevitably bring noise into the system. However, passive PT-symmetric acoustic systems could be constructed to avoid the additional noise due to the gain, as has been demonstrated for optical systems[26-28].

In conclusion, we demonstrate a compact piezoelectric platform on lithium niobate for non-Hermitian and nonreciprocal acoustics. We choose an operating frequency of 200 MHz for simplicity of electronic circuit design, but it can be adjusted from a hundred megahertz to a few gigahertz by geometrically scaling the design[39]. While our demonstration uses discrete electronic components, a fully integrated high-frequency acoustic platform can be built using application-specific integrated circuits. Our approach provides a feasible toolbox for manipulating, routing, and amplifying microwave phonons in integrated devices. Combining nonreciprocal SAW devices with other integrated optical and electrical devices could enable profound non-Hermitian phononics. Finally, our active SAW circuits could enrich signal processing for next-generation wireless communication.

**Figures**

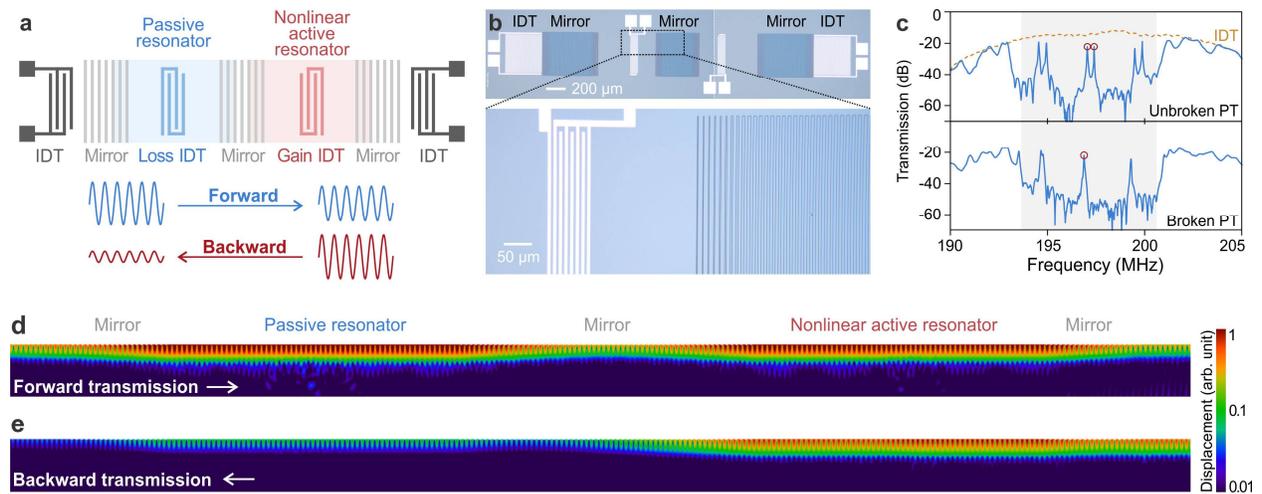

**Figure 1 | Nonreciprocal SAW transmission using nonlinear PT-symmetric resonators. a**, Schematic of our coupled SAW resonator system for nonreciprocal transmission. Loss and gain are introduced in the passive and active resonators, respectively. IDTs are used to create gain, loss and to generate and receive SAWs. For devices in the broken PT-symmetric regime, higher transmission is expected in the forward direction than the backward direction. **b**, Microscopic images of our fabricated device. The dark and bright areas are grooves etched into lithium niobate surface and aluminum electrodes, respectively. The large square aluminum pads are used to connect to external circuits by wire bonding. The white dashed line in the top image indicates the stitching boundary between two optical fields used to image the device. **c**, Measured transmission spectra of one device with a strong coupling in the unbroken PT-symmetric (reciprocal) regime and the other device with a weak coupling in the broken PT-symmetric (nonreciprocal) regime. Resonances are observed in the bandgap (highlighted in grey) of SAW Bragg mirror. For the coupled modes of interest (in red circles), a mode splitting is observed in the unbroken PT-symmetric regime, while only a single resonance is observed in the broken PT-symmetric regime. Transmission spectrum of a pair IDTs is also plotted. **d**, **e**, Numerical simulations of the magnitude of elastic displacement due to SAWs propagating through broken PT-symmetric (nonreciprocal) resonators in the forward (**d**) and backward (**e**) directions. The plots are stretched in the vertical direction for clarity.

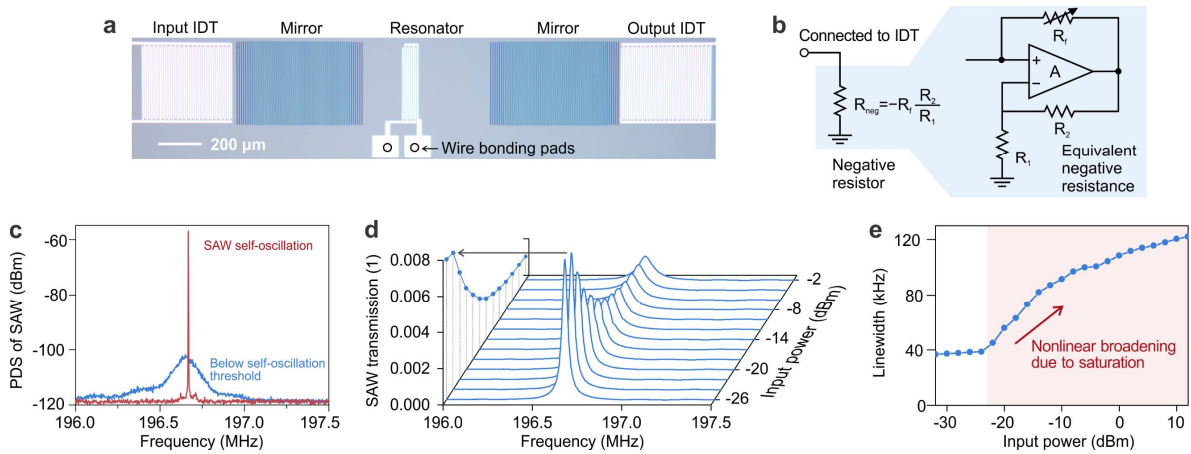

**Figure 2 | Characterization of SAW gain and nonlinearity. a**, Microscopic image of a single SAW resonator for gain and nonlinearity characterization. The gain for the SAWs is provided by the interdigital electrodes connected to an electronic circuit by wire bonding. **b**, Electronic diagrams of the negative resistor providing SAW gain. The negative resistance is implemented by an operational amplifier with resistive feedback. The equivalent negative resistance is tuned using the adjustable resistor $R_f$. **c**, Measured power spectral density (PSD) of the output SAW from the active resonator. When the gain is below the threshold for self-oscillation, the spectrum of the SAW collected at the output IDT shows a resonance with a linewidth of 94(1) kHz determined by a quality factor of the resonator (Q = 2,000). When the gain is increased above the threshold, resonator linewidth narrows below 2.5 kHz (the spectral resolution set by the spectral analyzer) and SAW self-oscillation is observed. **d**, Transmission spectra and **e**, linewidth of the nonlinear active resonator for various SAW input powers. The transmission resonance broadens, and its amplitude reduces for greater input powers due to the nonlinear saturation.

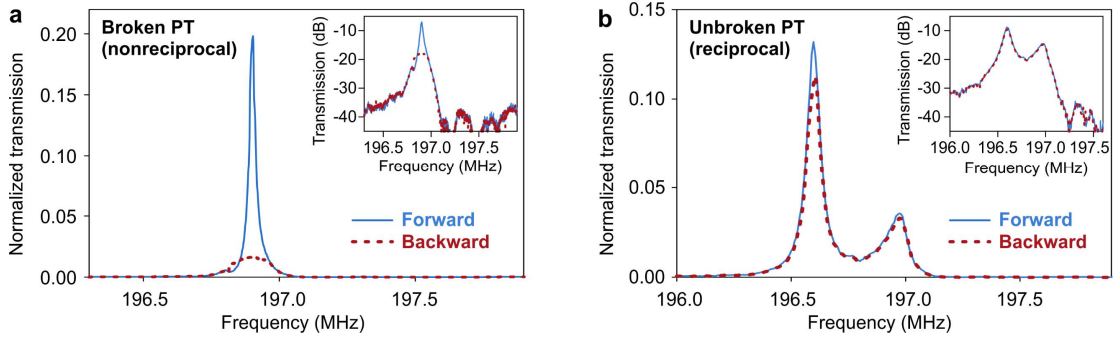

**Figure 3 | SAW Transmission of the nonlinear PT-symmetric resonators**. **a**, Nonreciprocal transmission measurements of the SAW resonators in the broken PT-symmetric regime. **b**, Reciprocal transmission measurements of the SAW resonators in the unbroken PT-symmetric regime. The microwave powers applied to input IDTs are -25 dBm (3 µW) in both plots. Insets: transmission plotted on a logarithmic scale. The transmission is normalized to that of a pair of IDTs.

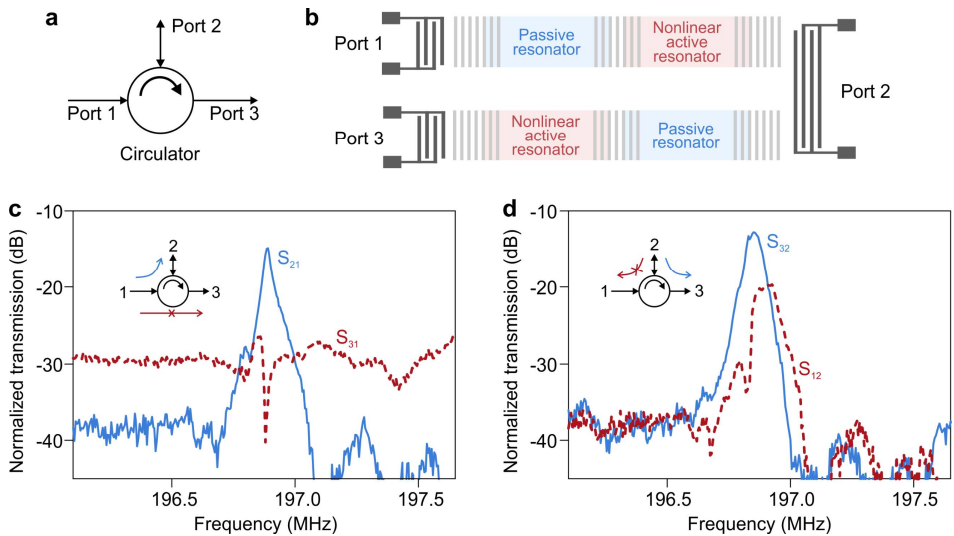

**Figure 4 | One-way circulation of SAWs. a**, Illustration of circulation that allows transmission in the clockwise direction, i.e. from port 1 to port 2, port 2 to port 3, and blocks transmission in the counterclockwise direction, i.e. from port 2 to port 1, port 3 to port 2. **b**, Schematic of the SAW circulation with two nonreciprocal devices connected in series. **c**, **d**, Transmission measurements between different ports. The transmission is normalized to that of a pair of IDTs.

## Methods

**Device design and fabrication.** Our SAW devices are fabricated on a 128° Y-cut LN substrate, and the SAW propagates in the X-direction of the crystal. The SAW resonators are defined by the Bragg mirrors, which are periodically etched grooves. The length of the single SAW resonator is 600 µm (between the edges of the Bragg mirrors). The grooves are patterned by photolithography using a maskless aligner (Heidelberg Instruments MLA 150), and the LN is etched by reactive-ion etching using Argon gas. The pitch of the grooves is 10 µm, which matches the half-wavelength of the SAW for a frequency of around 197 MHz. The etch depth is 0.5 µm. The width of the grooves is 5 µm at the center of the Bragg mirror and tapered to zero within ten periods at both ends of the Bragg mirror to reduce the acoustic loss, as shown in Fig. 1b. The number of grooves for the external Bragg mirror is 54. The number of grooves for the coupling Bragg mirror, which determines the coupling strength between two resonators, varies from 20 to 90 for strongly-coupled to weakly-coupled resonators. IDTs, electrical connections, and contact pads are fabricated using 120 nm thick aluminum deposited by thermal evaporation and patterned by photolithography and a lift-off process. The electrode pitch (width) of the IDT is 9.8 µm (4.9 µm), and its central frequency matches the SAW resonance frequency. The presence of the metal reduces the speed of the SAW, and thus the pitch of the IDT is slightly smaller than the pitch of the groove. The number of electrodes is 38 (19 signal and 19 ground electrodes) for the external IDTs outside the SAW resonators. These IDTs give a bandwidth of about 10 MHz from the transmission measurements and cover the reflection band of the Bragg mirror. The number of electrodes is 8 (4 signal and 4 ground electrodes) for IDTs inside the resonator that demonstrate SAW loss and gain with external electronic circuits. The contact pads are wire bonded to the printed circuit board (Fig. S8).

**Acknowledgement**
We thank Stefan Bogdanovic, Mengjie Yu, Mian Zhang, Cleaven Chia, Bartholomeus Machielse, and Yun-Feng Xiao for fruitful discussions.

**Funding:** This work is supported by the STC Center for Integrated Quantum Materials, NSF Grant No. DMR-1231319, NSF CQIS Grant No. ECCS-1810233, ONR MURI Grant No. N00014-15-1-2761, AFOSR MURI Grant No. FA9550-14-1-0389. N.S. acknowledges support by the Natural Sciences and Engineering Research Council of Canada (NSERC), the AQT Intelligent Quantum Networks and Technologies (INQNET) research program, and by the DOE/HEP QuantISED program grant, QCCFP (Quantum Communication Channels for Fundamental Physics), award number DE-SC0019219. This work was performed in part at the Center for Nanoscale Systems (CNS), Harvard University.

**Author contributions:** L.S. conceptualized, designed, fabricated, and measured the devices. W.M. and Y.H. theoretically analyzed the system with discussion from other authors. W.M. and L.S. performed numerical simulations. All authors analyzed and interpreted the results. L.S. and W.M. prepared the manuscript with contributions from all authors. M.L. and L.Y. supervised the project.

**Competing interests:** M.L. is involved in developing lithium niobate technologies at HyperLight Corporation. Other authors declare no conflict of interest.

**Data availability**: The datasets generated during and/or analyzed during the current study are available from the corresponding authors on reasonable request.

# Supplementary Text

## I. Construction of PT-symmetric Surface Acoustic Wave Resonators

Generally, non-Hermitian systems that respect parity-time (PT) symmetry can be constructed by introducing gain and loss in a spatially symmetric system. We construct a PT-symmetric surface acoustic wave (SAW) system using two resonators that feature loss and gain respectively, as shown in Fig. 1. The Hamiltonian of the system is given by

$$H = (\omega_1 + i\gamma)a_1^\dagger a_1 + (\omega_2 + ig)a_2^\dagger a_2 + \mu(a_1^\dagger a_2 + a_1 a_2^\dagger) \quad (S1)$$

in which $a_{1,2}$ are Bosonic annihilation operators for the SAW resonator 1 and 2, respectively, $\omega_{1,2}$ are the resonance frequencies of two resonators, and $\mu$ is the coupling strength between two resonators. The total loss of the first (passive) resonator is $\gamma < 0$, and the net gain of the second (active) resonator is $g > 0$. For two SAW resonators with same resonant frequency $\omega_0 = \omega_1 = \omega_2$, the eigenvalues of the two supermodes of the coupled system are given by

$$\omega_\pm = \omega_0 + i(\frac{\gamma+g}{2}) \pm \sqrt{\mu^2 - \mu_{PT}^2},$$

in which $\mu_{\text{PT}} = (g + |\gamma|)/2$ is the critical coupling strength also as known as the exceptional point for a PT system. The two supermodes, represented in the basis of individual resonant modes ($a_1$ and $a_2$), are given by

$$\begin{pmatrix} c_+ \\ c_- \end{pmatrix} = \begin{pmatrix} 1 & i\tilde{\mu}^{-1} + \sqrt{1 - \tilde{\mu}^{-2}} \\ i\tilde{\mu}^{-1} + \sqrt{1 - \tilde{\mu}^{-2}} & -1 \end{pmatrix} \begin{pmatrix} a_1 \\ a_2 \end{pmatrix},$$

in which $\tilde{\mu} = \mu/\mu_{\text{PT}}$ is the normalized coupling strength. When the coupling strength $\mu > \mu_{\text{PT}}$, the system is in the unbroken PT regime, and a mode splitting $[\text{Re}(\omega_+ - \omega_-) > 0]$ is expected. When $\mu < \mu_{\text{PT}}$, the system is in the broken PT regime, and the real parts of the eigenvalues of the two supermodes are same $[\text{Re}(\omega_+ - \omega_-) = 0]$.

Figure S1 shows the PT phase diagram of our coupled SAW resonators. In experiments, we first investigate the coupling strength between two passive resonators. The coupling strength is determined by the number of Bragg mirror periods between the two SAW resonators. In this case, two resonators feature the same loss ($\gamma = g < 0$), and mode splitting is observed. We introduce additional gain and loss for the SAW resonators by the in-resonator IDTs which are connected to external electronic circuits. In this case, when the coupling strength is smaller than the critical coupling strength $\mu_{\text{PT}}$, the system is in the broken PT-symmetry regime, and only a single resonant peak is observed in the transmission (Fig. 1c). For devices with increased coupling strength, our system enters the unbroken PT-symmetry regime, and mode splitting is observed. In experiments, to avoid self-oscillation of the electronic circuits, the loss of the SAW resonator is not completely compensated by the introduced gain. Our system is, however, equivalent to a time translation of a PT-symmetric system with balanced gain and loss, and it has features similar to that of a balanced PT system.

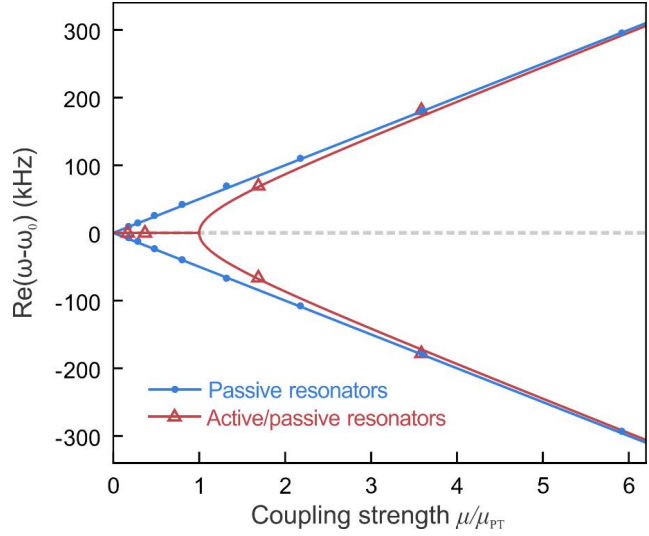

**Figure S1 PT phase diagram of coupled SAW resonators.** The lines are obtained from a theoretical calculation with parameters from experimental measurements. The data are obtained from the Lorentzian fitting of experimentally measured transmission spectra. The results for the case of two passive SAW resonators are shown in blue, and that for coupled active and passive resonators are shown in red.

## II. SAW transmission of the nonlinear PT-symmetric resonators

1. Forward and backward transmission

As discussed in the main text, the nonreciprocity of our PT-symmetric acoustic resonators relies on nonlinearity. We integrate the nonlinearity into the gain $g$ of the resonator. The total gain of the acoustic resonator is decomposed by $g = \frac{g_t}{2} - \frac{\gamma_0}{2} - \frac{\kappa_0}{2}$, where $\gamma_0$ and $\kappa_0$ are the intrinsic loss and the external coupling rate, respectively. The saturable nonlinear gain $g_t$ is given by $g_t = \frac{g_{t0}}{1+|a/a_0|^\beta}$, where $g_{t0} > 0$ is the gain in the weak intensity limit, $a$ is the amplitude in the resonator, $a_0$ and $\beta$ is the saturation amplitude and index describing the saturation behavior, respectively. The total loss for the passive resonator is given by $\gamma = -\frac{\gamma_e}{2} - \frac{\gamma_0}{2} - \frac{\kappa_0}{2}$, where $\gamma_e < 0$ is the additional acoustic loss induced by the electronics.

To calculate transmission spectra, we add input terms to the Hamiltonian (Equ. S1). The Hamiltonian with forward (labeled in $a$) and backward ($b$) inputs is given by

$$H_a = (\omega_1 + i\gamma)a_1^\dagger a_1 + (\omega_2 + ig_a)a_2^\dagger a_2 + \mu(a_1^\dagger a_2 + a_1 a_2^\dagger) + \epsilon_{\text{in}}(a_1 e^{i\omega t} + a_1^\dagger e^{-i\omega t})$$

$$H_b = (\omega_1 + i\gamma)b_1^\dagger b_1 + (\omega_2 + ig_b)b_2^\dagger b_2 + \mu(b_1^\dagger b_2 + b_1 b_2^\dagger) + \epsilon_{\text{in}}(b_2 e^{i\omega t} + b_2^\dagger e^{-i\omega t}),$$

where $\epsilon_{in} = \sqrt{\kappa_0 P_{in}/\hbar\omega}$ is the input strength into the resonator with SAW power $P_{in}$ at frequency $\omega$. The equations of motion for the system for forward and backward SAW transmissions are determined by the Heisenberg-Langevin Equation:

$$i\frac{d}{dt}\begin{pmatrix} a_1 \\ a_2 \end{pmatrix} = \begin{pmatrix} \omega_1 - \omega + i\gamma & \mu \\ \mu & \omega_2 - \omega + ig_a \end{pmatrix}\begin{pmatrix} a_1 \\ a_2 \end{pmatrix} + \begin{pmatrix} \epsilon_{in} \\ 0 \end{pmatrix}$$

$$i\frac{d}{dt}\begin{pmatrix} b_1 \\ b_2 \end{pmatrix} = \begin{pmatrix} \omega_1 - \omega + i\gamma & \mu \\ \mu & \omega_2 - \omega + ig_b \end{pmatrix}\begin{pmatrix} b_1 \\ b_2 \end{pmatrix} + \begin{pmatrix} 0 \\ \epsilon_{in} \end{pmatrix}.$$

These equations of motion are solved numerically to investigate the transmission. The forward and backward transmissions are given by

$$S_{Fwd} = \kappa_0 |a_2|^2 \hbar\omega / P_{in}$$

$$S_{Bwd} = \kappa_0 |b_1|^2 \hbar\omega / P_{in}.$$

Further, the nonreciprocal ratio $\eta$ is defined as

$$\eta = \frac{S_{Fwd}(f_0)}{S_{Bwd}(f_0)},$$

where $f_0$ is the frequency at which the forward transmission is maximized.

2. Strong coupling and reciprocal transmission

First, we consider the system with strong coupling between two resonators, in the unbroken-PT-symmetry regime, i.e. $\tilde{\mu} = \mu/\mu_{PT} > 1$. In this case, the real parts of eigenvalues $\omega_\pm$ bifurcate (mode split). In the strong coupling limit $\tilde{\mu} \to \infty$ the supermodes can be expressed as $c_\pm = \frac{(a_1 \pm a_2)}{\sqrt{2}}$, which corresponds to the coalescence of the two original modes. In this unbroken-PT-symmetry regime, the energy in the active resonator can flow into the passive resonator quickly enough to compensate the loss, leading to equilibrium. Therefore, the system with two strongly coupled resonators feature reciprocal transmissions. To validate this statement, we calculate the transmission in both directions, with the theoretically derived reciprocal transmission spectra shown in Figs. S2a and S2b. The system keeps its reciprocity with carried input power, and does not exhibit the power-related behavior of weakly coupled resonators (discussed in the next section). We further compare the mode amplitude in the active resonator between the forward and backward inputs (i.e. $a_2$ and $b_2$). As shown in Fig. S2c, the strong coupling ($\tilde{\mu} \gg 1$) results in $|a_2|/|b_2|$ tending to unity, which indicates that the system is of a similar state under forward and backward inputs. Reciprocity is thus expected for the strongly coupled SAW resonators.

Interestingly, we note that the normalized transmission for certain input powers (e.g. 1 μW and 100 μW in Figs. S2a and S2b) exceeds one. In fact, linear PT-symmetric systems, in which the gain is constant and not related to the intensity in resonators, can demonstrate spectral singularities, where the transmission becomes infinitely large[36]. Although the nonlinearity of our gain-loaded SAW system pulls the transmission singularity down to a finite value, the gain under weak input power can still result in a transmission that is greater than unity. When the system features a stronger input (e.g. 1 W in Figs. S2a and S2b), the saturated gain reduces the transmission intensity to less than one, resulting in an overall net loss.

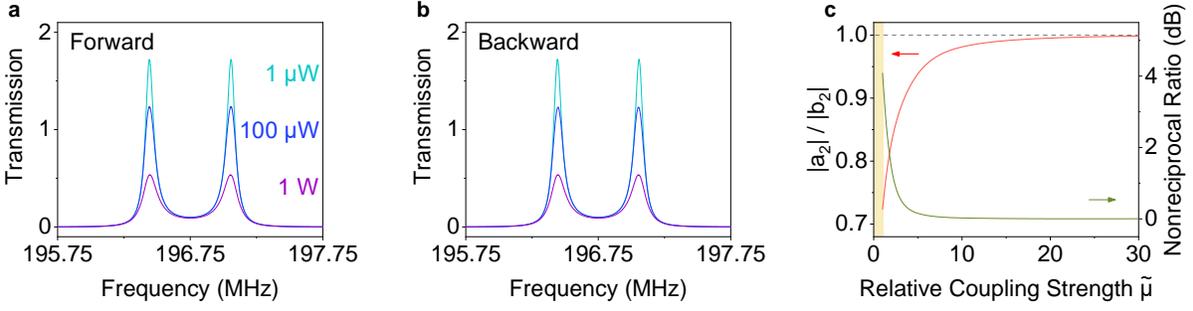

**Figure S2 Calculated transmission in the strong coupling region. a**, **b**, Forward and backward transmission in the strong coupling region $\tilde{\mu} = 10$ with varied input power (1 µW, 100 µW and 1 W). The SAW system with strongly coupled resonators has reciprocal transmission. The transmission at certain input power levels can be larger than one. **c**, The ratio of fields in the active resonator $|a_2|/|b_2|$ (red) and the nonreciprocal ratio (green) for different coupling strengths. The ratio of fields tends to one while the nonreciprocal ratio tends to zero. The input power is 100 µW, and the highlighted area represents the weak coupling region.

3. Weak coupling and nonreciprocal transmission

The system enters the broken-PT symmetry regime when the coupling strength reduces to less than the PT-symmetry transition point (i.e. the exceptional point), $\tilde{\mu} < 1$. We note the difference from the artificially broken symmetry, the broken-PT-symmetry phase here is caused by the weaker coupling strength between the two resonators. Weakly-coupled resonators nearly behave as their own independent states – the supermodes $c_+$ and $c_-$ share the same real eigenvalues but the opposite signs of the imaginary parts. In this case, the energy flow between two resonators is not quick enough, and the system is not in equilibrium. Nonreciprocal transmission spectra are numerically calculated under a 100 µW input (Fig. S3b). We note the cause of the nonreciprocal transmission spectra is the induced unequal nonlinear gains $g_a \neq g_b$ for the forward and backward inputs by symmetry breaking.

The nonreciprocal ratio is related to the input power, shown in Fig. S3d. For weak SAW input powers, the gain is under the saturation threshold and almost linear (Fig. S3c), and the transmission spectra are almost reciprocal, as shown in Fig. S3a. For higher input powers, the gain of the active resonator is significantly lower for the backward input than the forward input due to the saturation (Fig. S3c). In this case, our calculations yield strongly nonreciprocal transmission spectra as shown in Fig. S3b.

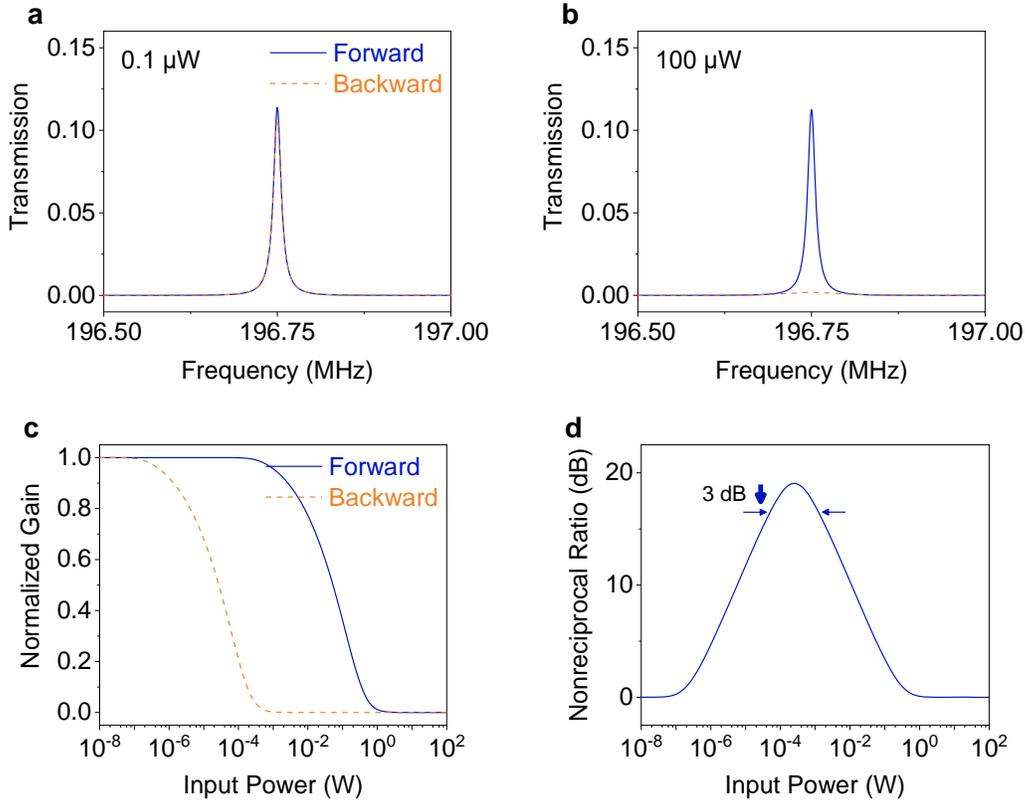

**Figure S3 Calculated transmission, saturable gain, and nonreciprocal ratio in the weak coupling regime. a**, **b**, Normalized transmission with an input power 0.1 µW and 100 µW, respectively. **c**, External gain for forward and backward propagation, normalized to the value with a weak input power. The gain is suppressed at high input powers due to the nonlinear saturation. **d**, Nonreciprocal ratios with different input powers. The dynamic range (labeled by two blue arrows) is defined as the range of input power where the nonreciprocal ratio decreases by 3 dB. In (**a**)-(**d**), the relative coupling strength $\tilde{\mu} = 0.05$, the additional loss induced by the electronics $\gamma_e = 0.2\gamma_0$, the initial external nonlinear gain $g_{t0} = 3.6\gamma_0$, and the external coupling rate $\kappa_0 = 3\gamma_0$, where $\gamma_0$ is the intrinsic loss of single resonator.

We investigate the nonreciprocal ratios under various input powers and gains. The nonreciprocal ratios maximize at moderate inputs Fig. S3d, where the saturable gains also exhibit the largest difference between forward and backward inputs (Fig. S3c). At the weak input limit, the amplitudes in the resonators are far below the saturation threshold, and this linear system does not break the law of reciprocity. At the strong input limit, the SAW amplitudes in the resonators are far above the saturation, and the gain is negligible compared to the input powers. Thus, the system falls back into the reciprocal condition. Thanks to the tunable and saturable external gain, our SAW devices allow us to study the different regimes of a nonlinear PT-symmetric system.

## III. Performance of the nonreciprocal SAW device

We evaluate the performance of our nonreciprocal device including nonreciprocal ratio, dynamic range, and insertion loss as the system parameters are varied.

### 1. Nonreciprocal ratio

The nonreciprocal ratios at the optimal input power depend on the relative coupling strengths $\tilde{\mu}$ and the small signal gains $g_{t0}$ (Fig. S4a). A weaker coupling strength leads to a higher nonreciprocal ratio. The weaker coupling strength, however, increases the insertion loss. Alternatively, a larger gain in the active resonator improves the nonreciprocal ratios as well. Theoretically, a nonreciprocal ratio of over 35 dB could be achieved at the gain $g_{t0} = 3.95\gamma_0$ (Fig. S4a). However, in our experiments, the gain is limited by the operational amplifier used in the external electronic circuits, and a 10-dB nonreciprocal ratio is experimentally demonstrated.

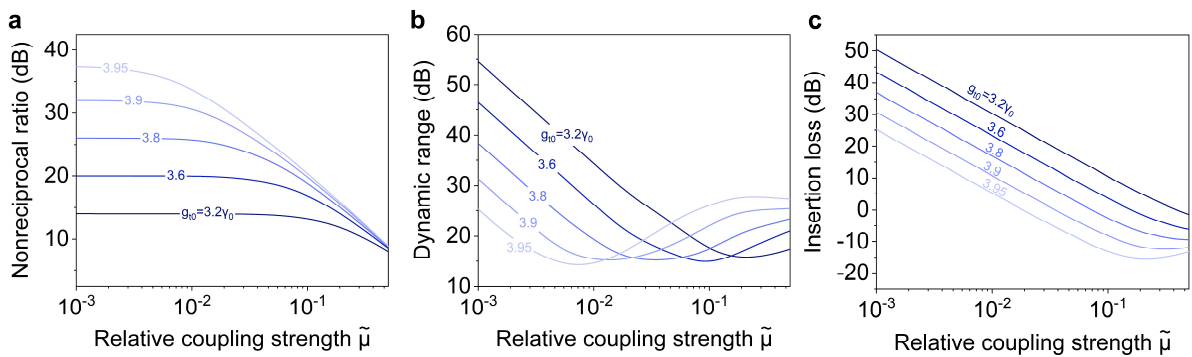

**Figure S4 Performance of nonreciprocal SAW devices. a**, Nonreciprocal ratio using optimal input power. **b**, Dynamic range within 3-dB reduction of the nonreciprocal ratio, as illustrated in Fig. S3d. **c**, Insertion loss of the nonreciprocal devices only accounting for the coupled resonators, and IDT insertion loss is not included. Results are derived from the numerical simulation of the system dynamics (refer to Sec. II of this Supplementary) for varying coupling strength $\tilde{\mu}$ and gain $g_{t0}$. The gain $g_{t0}$ is normalized to the intrinsic resonator loss $\gamma_0$, and varies from $3.2\gamma_0$ to $3.95\gamma_0$ as shown in gradient blue colors. Other parameters are as same as in Fig. S3.

### 2. Dynamic range

The dynamic range says how much the input power can vary without significant performance degradation. The working input powers result in nonreciprocal ratios within 3-dB degradation from the optimum (Fig. S3d), and the dynamic range is defined as the ratio between the maximum and minimum working input powers. A dynamic range of more than 40 dB is theoretically possible for our nonreciprocal acoustic system (Fig. S4b). In the weak coupling limit, a lower gain in the active SAW resonator leads to a higher dynamic range, but it degrades the nonreciprocal ratio. System parameters could be designed to reach a trade-off between the different performance parameters.

## 3. Insertion loss

In addition, we derive the insertion loss from the dynamics of our nonreciprocal SAW devices. Assuming the device is in the nonreciprocal weak coupling region ($\tilde{\mu} = \mu/\mu_{\text{PT}} < 1$) and the input power is small enough to assume the gain $g_a$ as a constant (i.e. in the linear region), the forward intensity transmission of our device is given by

$$S_{\text{Fwd}} = \frac{\kappa_0^2}{\left(\frac{|\gamma|g_a}{\tilde{\mu}\,\mu_{\text{PT}}} - \tilde{\mu}\,\mu_{\text{PT}}\right)^2},$$

where $\mu_{\text{PT}} = (g_a + |\gamma|)/2$. The insertion loss is given by $IL = -10\log S_{\text{Fwd}}$. A greater external coupling strength $\kappa_0$ will result in larger transmission (less insertion loss). However, it will reduce the loaded $Q$ of our SAW resonator and degrade the nonreciprocity. We note that the nonlinear gain $g_a$ depends on the input power and shifts the PT coupling strength $\mu_{PT}$. We numerically calculate the insertion loss for various coupling strengths and gains (Fig. S4c). Since our nonreciprocal device is active, the transmission can exceed one under certain relatively large coupling strength $\tilde{\mu}$ and results in a negative insertion loss. At the weak coupling strength $\tilde{\mu}$ limit, a greater gain will reduce the insertion loss.

In summary, there is a tradeoff between various performance parameters including nonreciprocal ratio, dynamic range and insertion loss. Generally, a weaker coupling strength between the resonators will lead to a larger dynamic range, but a greater insertion loss. A higher gain of the active resonator will result in a stronger nonreciprocity, less insertion loss, but a smaller dynamic range.

## IV. Numerical simulation of PT-symmetric SAW resonators

Our finite element simulation using COMSOL Multiphysics is based on the 2D cross-section of our device, and the dimensions of the structures are same as the actual device design (Fig. S5). The SAWs propagate in the crystal X direction of 128°Y-cut lithium-niobate, and we define the geometric x-axis to be parallel to the direction of propagation while the y-axis is perpendicular to the device surface. We employ three simulation modules: solid mechanics, electrostatics, and electronic circuits. The strain and electrostatic fields are coupled together via the piezoelectric effects. The bottom boundary of lithium niobate is a perfect matching layer with a fixed position condition in solid mechanics at the end, while the top surface is free. Two ends along the x-axis directions of our system are also set as the perfect matching layers to absorb out-propagating waves and avoid artificial reflections of acoustic waves. The IDT electrodes outside the SAW resonators are used for transmission measurements. The IDT electrodes inside the SAW resonators are connected to electronic circuits providing loss and nonlinear gain. We simulate the negative resistor by a voltage-controlled current source in the COMSOL circuit module, providing an equivalent nonlinear gain, expressed as $R_0\left(1 + |u/u_0|^\beta\right)$. Here, $R_0 < 0$ and $\beta$ is the index of the saturation rate as previously discussed, $u$ is the voltage over the negative resistor and $u_0$ is the threshold of saturation. This negative resistor is connected to the electrodes of the IDTs to provide external gain. A linear electronic resistor is connected to the IDTs in the passive resonator to provide loss.

For the weak coupling (broken-PT-symmetric) regime, the number of coupling mirrors is 80. The simulated nonreciprocal SAW transmission spectra are shown in Fig. S6a. For the forward SAW input, the acoustic fields are distributed over both SAW resonators, allowing for a higher transmission (Fig. S6b). For the backward SAW input, the active resonator has a lower gain due to the localization, resulting in a lower transmission (Fig. S6c).

For the strong coupling (unbroken-PT-symmetric) regime, the number of coupling mirrors is reduced to 20. Mode splitting and reciprocal transmission spectra are shown in Fig. S7a. Similar amplitudes of SAW between the active and passive resonators are observed for both forward and backward SAW inputs, and this results in a reciprocal phonon transmission (Figs. S7b and S7c).

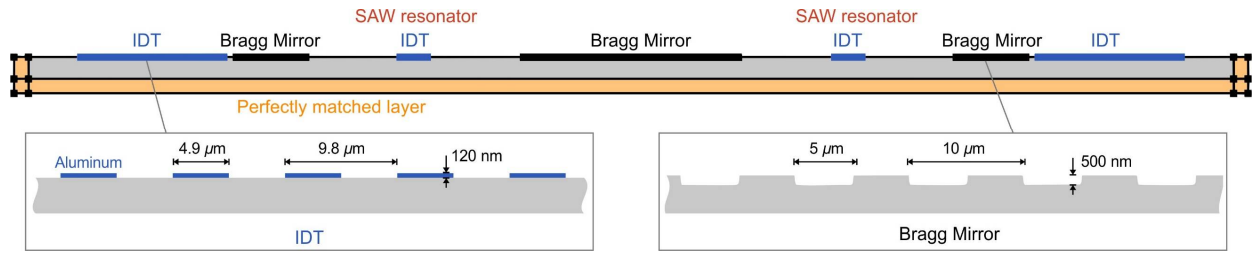

**Figure S5 Device configuration for the numerical simulation.** The SAW resonators are formed between the Bragg mirrors. The IDTs inside the resonators are connected to electronic circuits that provide loss and gain. The IDTs on the ends are used to measure the SAW transmission.

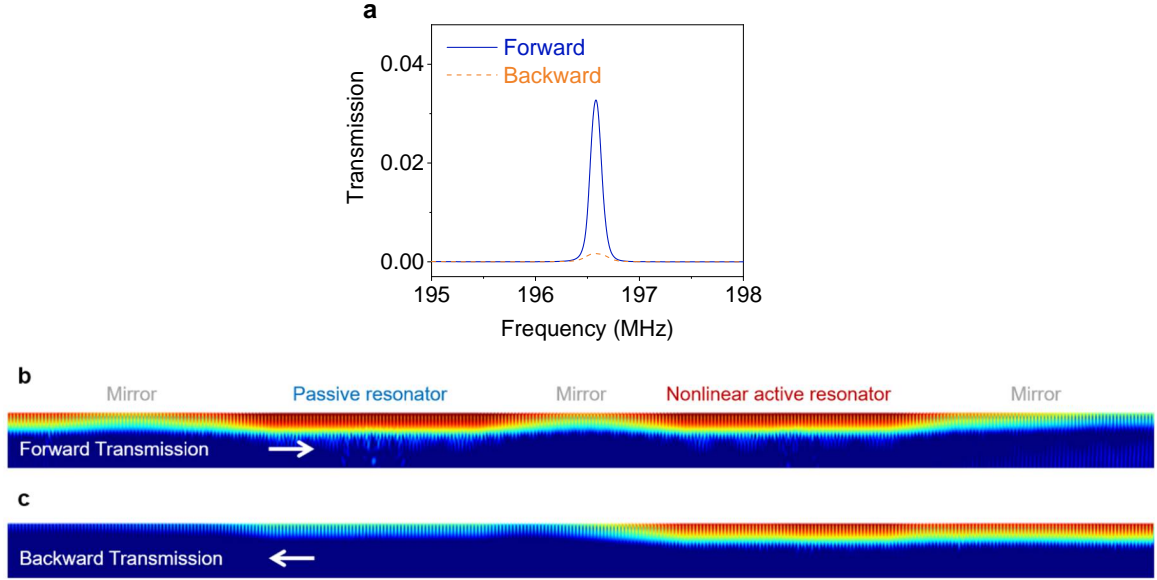

**Figure S6 Numerical simulation of nonreciprocal transmission in the weak coupling regime. a**, Simulated nonreciprocal transmission of the weakly coupled PT-symmetric SAW resonators. **b**, **c**, Simulated SAW profiles for the forward and backward inputs, respectively. The SAW fields with forward input distribute over both SAW resonators while the mode mainly localizes in the active resonator for the backward input. **b**, **c**, replots the results in Figs. 1**d** and 1**e** for comparison with Fig. S7.

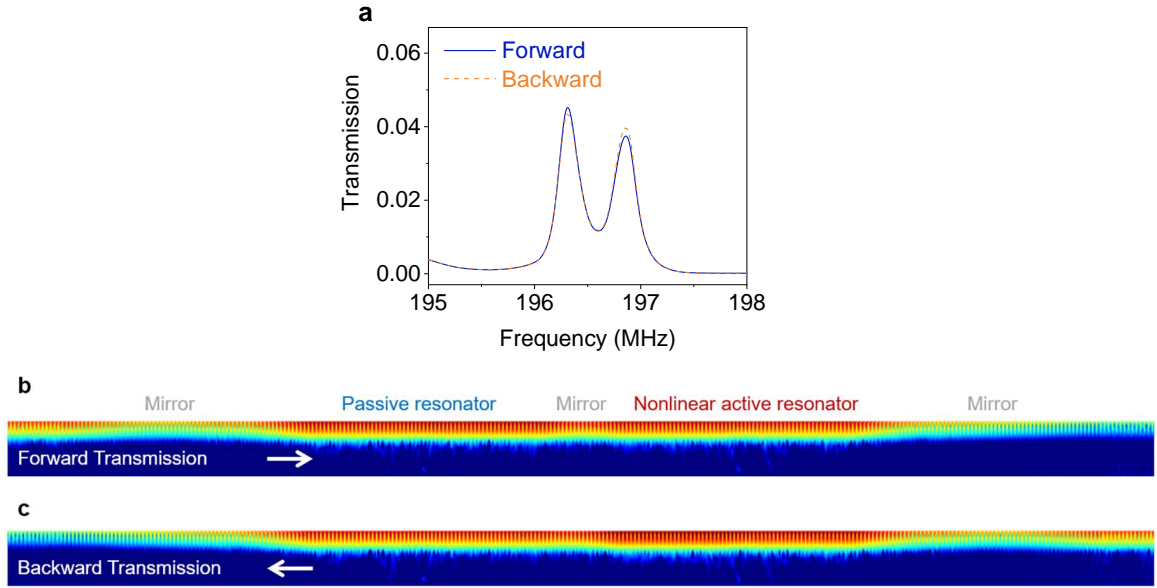

**Figure S7 Numerical simulation of PT-symmetric SAW resonators in the strong coupling regime.**
**a**, Simulated reciprocal transmission of the strongly coupled PT-symmetric SAW resonators. **b**, **c**, Simulated SAW profiles for the forward and backward inputs, respectively. Similar amplitudes of the active and passive SAW resonators are observed for both forward and backward SAW inputs.

### V. Electronic circuits used for loss and nonlinear saturable gain

Electronic circuits are connected to distributed electrodes (i.e. IDTs) on the surface of lithium niobate to achieve gain and loss for the SAWs. Here, we describe the implementation of the electronic circuits on a printed circuit board (PCB) (Fig. S8). The lithium niobate SAW chip with coupled resonators is placed at the center of the PCB. The electrodes on chip are wire bonded to the contact pads on the PCB to allow transmission measurements and tailoring of the electrical fields of each SAW. The electronic schematic for the negative resistance and the regular resistance circuits are shown in Fig. S9. The resistance circuit, connected to the passive SAW resonator, consists only of a trimmer resistor. The equivalent negative resistance circuit, connected to the active SAW resonator, consists of an operational amplifier (TI LMH6629), resistors, and capacitors. The circuit is designed to separate DC and AC signals by coupling capacitors $C_f$ and $C_1$. For DC bias, the resistors $R_3$ and $R_4$ provide the input bias of VCC/2 at the positive input of the amplifier; and the closed loop amplification is unity gain. For AC signals, the effective negative resistance at the connector is $R_{neg}=-R_f R_2/R_1$. The trimmer resistor $R_f$ is tunable from 0 to 5 kΩ. We use $C_f$ = 500 pF, $C_1$ = 10 nF, $R_1$ = 100 Ω, $R_2$ = 1.5 kΩ, and $R_3 = R_4$ = 20 kΩ. The power supply VCC is 2.7 V. The saturation behaviors of the circuits are shown in Fig. S8. For the negative resistance circuit, the output voltage operational amplifier will be clamped by the power supply voltage VCC, and thus the reflected power shows a saturation for higher input powers. The standard electronic resistor that is used as the passive resistor behaves linearly within our range of input powers.

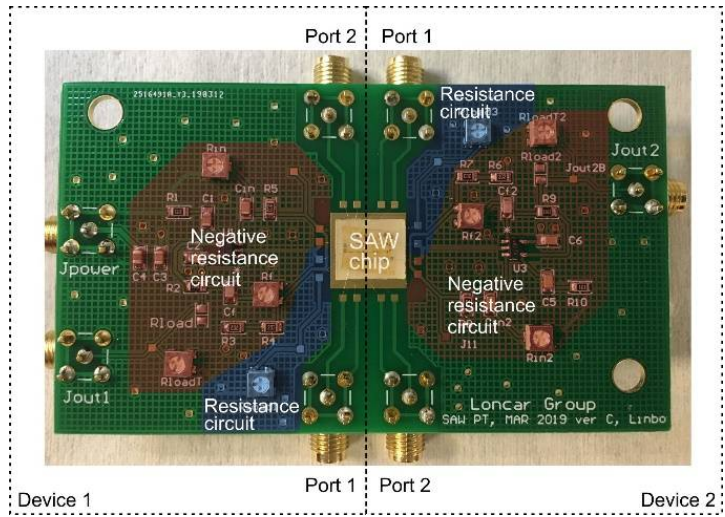

**Figure S8 Photo of the PCB used for nonreciprocal SAW device measurements.** This PCB supports simultaneous measurements of two devices. Signal from Port 1 to Port 2 is defined as the forward direction.

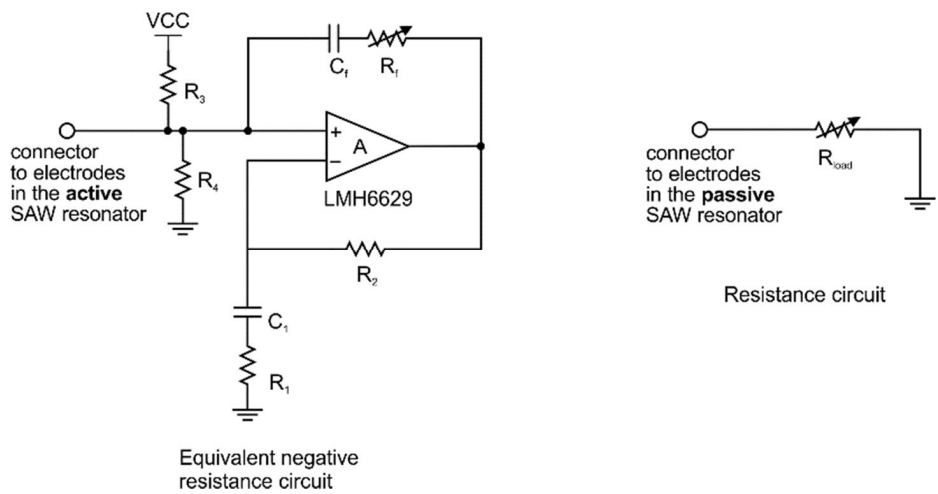

**Figure S9** Electronic schematic for circuits providing gain and loss of active and passive SAW resonators.

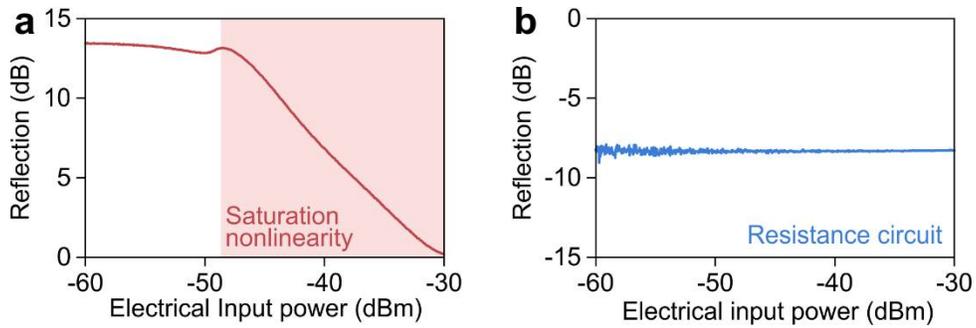

**Figure S10 Measured power reflectance of the circuits connected to active and passive SAW resonators. a**, Power reflection from the negative resistance circuit connected to the active SAW resonator. Reflection above 0 dB indicates a gain from the circuit, and saturation occurs at high input powers. **b**, Power reflection from an electronic resistor providing external loss of the passive SAW resonator. It is a linear circuit over our input power range.

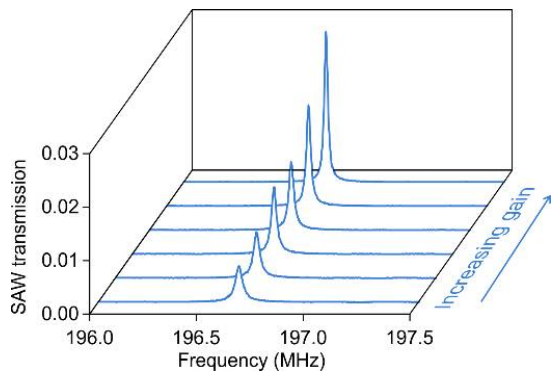

**Figure S11** Measured transmission of the active SAW resonator depicted in Fig. 2a for various gain values, below the SAW self-oscillation threshold.

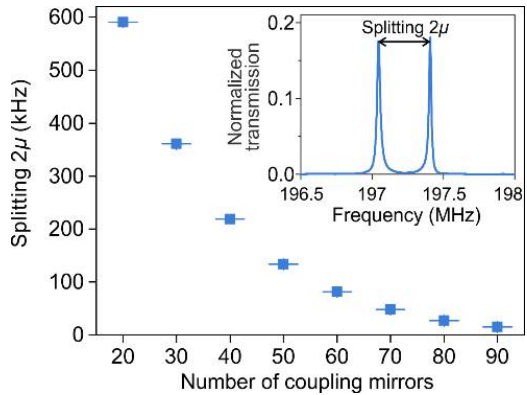

**Figure S12** Measured mode splitting of two coupled SAW resonators with varying numbers of Bragg mirror periods. Inset: the transmission of strongly coupled SAW resonators with coupling Bragg mirror of 30 periods.

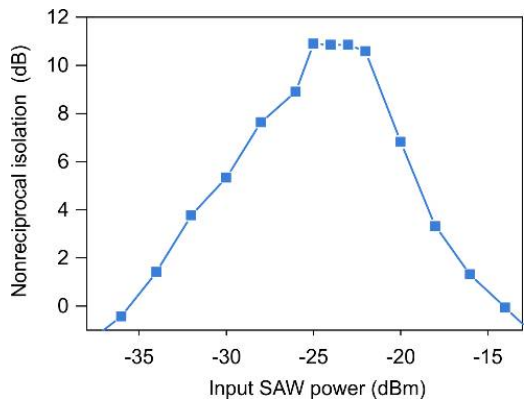

**Figure S13** Measured nonreciprocal isolation of broken-PT-symmetric SAW resonators.

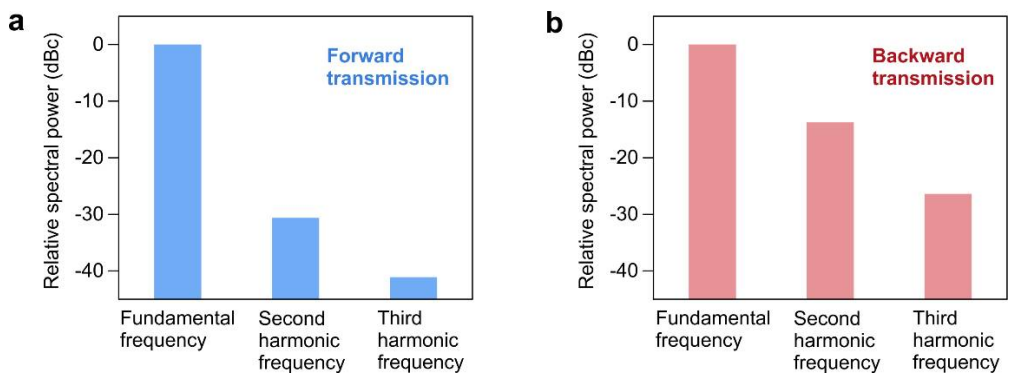

**Figure S14** Measured power at higher harmonic frequencies for (**a**) forward and (**b**) backward transmitted SAWs using the nonreciprocal device. All powers are referenced to the transmitted power at the fundamental frequency.